\documentclass[preprint,showpacs,preprintnumbers,amsmath,amssymb,aps]{revtex4}

\usepackage{graphicx}
\usepackage{dcolumn}
\usepackage{bm}

\begin{document}

\title{Energy dependence of the single spin asymmetries 
       in inclusive pion production processes}
\author{Dong Hui}\email{donghui@mail.sdu.edu.cn}
\author{Li Fang-zhen}
\altaffiliation[Now at ]{Institute of Biophysics, 
Chinese Academy of Science, Beijing, China}
\author{Liang Zuo-tang}\email{liang@sdu.edu.cn}
\affiliation{Department of Physics,Shandong University, 
Jinan, Shandong 250100, China}

\begin{abstract}
Recent data from E925 Collaboration at Brookheaven National Laboratory 
show a significant energy dependence of the single-spin 
left-right asymmetry in inclusive hadron production in 
hadron-hadron collisions. 
We analyzed the experimental results and show that 
the observed energy dependence can be reproduced 
naturally in the picture proposed in a previous Letter. 
We fixed all the parameters at the Fermilab E704 energy 
and calculate the asymmetry at the BNL E925 energy. 
We compare the results with the data and make predictions 
for experiments at even higher energies.  
\end{abstract}

\maketitle

The existence of striking single-spin 
left-right asymmetries ($A_N$) in inclusive 
hadron production in hadron-hadron collisions 
with transversely polarized beam 
has attracted much attention in the past years. 
In contrast to the leading order 
theoretical prediction made\cite{Kane78} in 1978, 
the data\cite{ANexp} show that 
the produced hadrons exhibit left-right asymmetries 
up to 40\% in the fragmentation region. 
A number of theoretical 
approaches\cite{Siv90,Qiu91,LM92,BLM93,LM94,Col93,Ans95,Efr95,Bro02,Boer99,Boer03,Bur02,LB00}
have been made, 
the aim of them is to explain why such striking effects exist. 
But the mechanisms which lead to such asymmetries are still in debate. 
New measurements at high energies are underway. 
It is interesting to note the recent data from 
the E925 Collaboration at Brookheaven National Laboratory (BNL)
for pion production in $pp$ collisions\cite{E925} at $p_{inc}=22$ GeV/c.
Compared with the results obtained earlier by 
Fermilab E704 Collaboration\cite{E704} at $p_{inc}=200$ GeV/c, 
the data from BNL E925 show the following characteristics: 
First, the E925 data confirms the striking features 
of the E704 data: 
(a) the magnitudes of $A_N$ are approximately zero for small $x_F$, 
they increase monotonically with increasing $x_F$ 
and reach about 40\% at $x_F\approx 0.8$; 
(Here, $x_F\equiv 2p_\|/\sqrt{s}$, 
$p_\|$ is the component of the momentum of the produced meson 
in the center of mass frame of the colliding hadron system parallel 
to the direction of motion of the incident hadron; 
$s$ is the center of mass energy squared.) 
(b) the sign of $A_N$ for $\pi^+$ in $pp$ collisions is positive 
and that for $\pi^-$ is negative. 
Second, at the same time, the E925 data show 
also the following striking energy dependence: 
The $|A_N|$'s at the E925 energy 
start to rise much later than those at the E704 energy. 
They start to rise at about $x_F\approx 0.5$ while 
those at the E704 energy start at $x_F\approx 0.3$.
It should be helpful in distinguishing 
different mechanisms to see whether such an energy 
dependence can be reproduced. 

One of the theoretical approaches to explain the asymmetries 
in the last years has been proposed by 
the Berliner group\cite{LM92,BLM93,LM94,LB00}.
The major characteristics of the picture are: 
(1) The produced mesons are divided into two categories: 
the ``direct-formation'' or ``direct-fusion'' part and the rest, 
where the former denotes those mesons which are directly produced 
and contain the valence quarks from the incident hadron. 
It has been shown that the former can be described by the 
direct-fusion process $q_v+\bar q_s\to M$ and contributes 
to the single-spin left-right asymmetry but the latter does not. 
(2) The left-right asymmetry for a meson 
from $q_v(\uparrow)+\bar q_s\to M$ arises because of 
the orbital angular momentum of the valence quark in 
the polarized proton and the ``surface effect'' during the 
hadron-hadron collisions. 
The picture has been applied to different processes
at E704 energy.
A good agreement with the various available data 
has been obtained with only two adjustable parameters\cite{LB00}. 
It should be interesting to see whether this picture can 
give rise to the energy dependence observed by E925 
Collaboration compared to the E704 data. 

In this note we study this problem and show that the 
energy dependence can easily be reproduced by the 
picture in \cite{BLM93} without adjusting or introducing 
any further parameters. 
We used the values of the two parameters fixed at 
the E704 energy to do the calculations at the 
E925 energy. We compare the results with the 
data and extend the discussion to even higher energy region.

We now start the discussion 
with a brief summary of the formulation 
of the picture in \cite{BLM93}. 
We use the same notations as in \cite{BLM93} 
and denote the number densities of the mesons 
from the direct-fusion
process $q_v+\bar q_s\to M$ and 
the rest by $D(x_F,s|M)$ and $N_0(x_F,s)$ respectively. 
Empirical facts 
(see the references given in \cite{BLM93,LM94,LB00})
show that $D(x_F,s|M)$ can be obtained from 
the quark distributions 
$q_v(x_v)$ and $\bar q_s(x_s)$ 
of the corresponding flavors. 
(Here, $x_v$ and $x_s$ are the momentum fractions carried by 
$q_v$ and $\bar q_s$ respectively.) 
Since energy-momentum conservation in the process 
requires that $x_v\approx x_F$ and $x_s\approx x_0/x_F$ 
(where $x_0 = m^2/s$, $m$ is the mass of the produced meson), 
it has been obtained that, 
\begin{equation}
D(x_F,s|M)=\kappa_M q_v(x_F)\bar q_s(x_0/x_F),
\end{equation}
where $\kappa_M$ is a constant which is fixed 
by one data point on the cross-section 
in the large $x_F$ region of the unpolarized reaction\cite{ft_kappa}.  
More precisely, for $\pi^+$ and $\pi^-$, 
we have, 
\begin{equation}
D(x_F,\pi^+|s)=\kappa_\pi u_v(x_F)\bar d_s(x_0/x_F),
\end{equation}
\begin{equation}
D(x_F,\pi^-|s)=\kappa_\pi d_v(x_F)\bar u_s(x_0/x_F).
\end{equation}
It should be emphasized that 
$D(x_F,M|s)$ dominates the large $x_F$ region, 
while the non-direct-formation part $N_0(x_F,M|s)$
comes mainly from the interaction of the seas 
(sea quarks, sea anti-quarks and gluons) of the two colliding 
hadrons and is dominant in small $x_F$ region.

In reactions with transversely polarized beams, 
the single-spin left-right asymmetry $A_N$ comes from 
the direct-formation part, and is given by\cite{BLM93,LB00}, 
\begin{equation}
A_N(x_F,M|s)=
\frac{C\Delta D(x_F,M|s)}{N_0(x_F,M|s)+D(x_F,M|s)},
\label{eq.define}
\end{equation}
where $\Delta D(x_F,M|s)=\kappa_M \Delta q_v(x_F|tr)\bar q_s(x_0/x_F)$
is the difference between the number density
of $M$ from $q_v+\bar q_s\to M$ 
where the $q_v$ is polarized in the same 
and that in the opposite direction as the proton   
in reaction with upward polarized proton beam; 
$\Delta q_v(x_F|tr)=q_v^+(x_F,\uparrow\nobreak)-q_v^-(x_F,\uparrow)$ 
and $q_v^\pm (x_F,\uparrow)$ is the number density of $q_v$ 
in a transversely polarized proton where the polarization 
of $q_v$ is the same ($+$) as or opposite ($-$) to the proton.  
We note that the $\Delta q_v(x|tr)$ introduced in \cite{BLM93} is nothing 
else but the transversity distribution $\delta q_v(x)$ now 
discussed frequently in the literature. 
The constant $C$ is the difference between 
the probability for $M$ produced in $q_v+\bar q_s\to M$ 
to go left and that to go right if the $q_v$ is upward polarized. 
It is the second parameter in the model and 
is fixed\cite{BLM93} to be $C=0.6$ by one $A_N$ data 
in the large $x_F$ region. 

As has been emphasized above, 
for large $x_F$, $D(x_F,M|s)\gg N_0(x_F.M|s)$. 
Hence, 
\begin{equation}
A_N(x_F,M|s)\to \frac{C\Delta D(x_F,M|s)}{ D(x_F,M|s)}
=\frac{C\Delta q_v(x_F|tr)}{q_v(x_F)}, \ {\rm at\ } x_F\to 1.
\end{equation}
For small $x_F$, $D(x_F,M|s)\ll N_0(x_F.M|s)$. 
Hence, 
\begin{equation}
A_N(x_F,M|s)\to 0, \ {\rm at\ } x_F\to 0,
\end{equation}
This implies that for $x_F$ goes from zero to one, 
$|A_N|$ should start from zero, begin to rise somewhere 
and reach $C\Delta q_v(x_F|tr)/q_v(x_F)$ at large $x_F$.
We also clearly see that where $|A_N|$ begins to rise is determined by 
the interplay of the two contributions $D(x_F,M|s)$ and $N_0(x_F,M|s)$. 
Whether it is dependent of energy is determined by the energy 
dependences of $D(x_F,M|s)$ and $N_0(x_F,M|s)$.

Eq.(\ref{eq.define}) has been applied 
to calculate the $A_N$'s for different mesons in $pp$ and 
$\bar pp$ collisions at the E704 energy 
and the results are in agreement with the data. 
Now we use it to discuss the energy dependence of
$A_N(x_F,M|s)$. 
We note that due to the scaling behavior of the inclusive 
cross section in hadron-hadron collisions 
in the central rapidity region\cite{Scaling}, 
we expected that $N_0(x_F|s)$ is independent of energy. 
The two parameters $C$ and $\kappa$ are two constants and 
are taken as independent of energy. 
There is an obvious energy dependence in $D(x_F,M|s)$ 
which is caused by the energy dependence of $x_0$. 
This leads to an energy dependence of $A_N(x_F,M|s)$ 
in the picture. 
We now analysis whether this energy dependence 
is in the right direction as 
that observed in experiments.

As can be seen from the expression of $x_0(=m^2/s)$, 
if $\sqrt{s}$ is higher, $x_0$ is smaller. 
Since in the small $x$ region 
$\bar q_s(x)$ increases rapidly with decreasing $x$, 
this leads to a rapid increase of 
$\bar q_s(x_s)$ (where $x_s\approx x_0/x_F$) 
with increasing $\sqrt{s}$. 
This means that, at higher $\sqrt{s}$, 
the number of $\bar q_s$ at $x_s$ is larger so that 
the probability for the $q_v$ to meet a suitable $\bar q_s$  
to form the meson $M$ is larger. 
A direct consequence is 
that $D(x_F,M|s)$ becomes more important 
compared to $N_0(x_F,M|s)$. 
Hence $|A_N|$ should starts to become non-zero 
earlier when $x_F$ increases from zero to one 
at higher $\sqrt{s}$. 
This qualitative expectation is in agreement with the 
experimental observation. 
Now, we use completely the same inputs as 
that used in \cite{BLM93} for $p_{inc}=200$ GeV/c 
but change the energy $\sqrt{s}$ to the E925 energy, 
i.e. take $\sqrt{s}=6.56$ GeV to calculate $A_N$ 
from Eq.(\ref{eq.define}).
The results obtained are given in Fig.1. 
We see that the qualitative feature of the data can indeed
be reproduced \cite{ft_pT}. 
The quality of the fit to the data depends 
on the choice of $\Delta q_v(x|tr)$. 
Since it is still unknown yet, in \cite{BLM93}, 
a simple ansatz $\Delta q_v(x|tr)\propto q_v(x)$ 
was used\cite{ft_pdf},
where the proportional constants were determined by 
using the SU(6) wavefunctions, i.e., 
$\Delta u_v(x|tr)=(2/3)u_v(x)$ 
and $\Delta d_v(x|tr)=(-1/3)d_v(x)$. 
To see whether the quality of 
the fit at the E925 energy 
and that at E704 energy can be improved simultaneously 
by making a more suitable choice of $\Delta q_v(x|tr)$, 
we take a more sophisticated form of $\Delta q_v(x|tr)$ 
to make a better fit of the E704 data 
and then apply to the E925 energy. 
More precisely, 
instead of $\Delta q_v(x|tr)\propto q_v(x)$, 
we introduce an extra $x$-dependent factor $f(x)$, 
i.e. take\cite{ft_transversity} 
$\Delta q_v(x|tr)\propto f_v(x) q_v(x)$,  
where $f_u(x)=1.2x$, $f_d(x)=2.6x^{1.5}$ 
[more explicitly, 
we take $\Delta u_v(x|tr)=0.8xu_v(x)$ 
and $\Delta d_v(x|tr)=-0.87x^{1.5}d_v(x)$]. 
The results are shown in Fig.2. 
We see that the quality of the fit at both energies 
can indeed be improved simultaneously. 

We see that without any further input or adjusting any parameter, 
we can reproduce the energy dependence observed by E925 compared 
with the E704 results. 
It reflects the energy dependence of the interplay 
of the direct-formation part and the rest. 
Encouraged by this result, we apply the picture to
make predictions for experiments at even higher 
energies, much higher than the E704 energy, such 
as those at RHIC.  
To do this, we recall that 
the energy dependence of $A_N$ in the picture\cite{BLM93} 
comes from that of $D(x_F,M|s)$. 
The latter depends on $\sqrt{s}$ because, 
when $\sqrt{s}$ increases, $x_0$ is smaller so that 
$\bar q_s(x_s)$ is larger and there is more 
$\bar q_s$'s suitable to combine with the $q_v$ to form the meson $M$. 
Since the number of $q_v$'s in a proton is very much limited, 
it can easily be imagined that,  
at very high $\sqrt{s}$, $\bar q_s(x_s)$ is very large  
so that there is always enough suitable $q_s$ to 
combine with $q_v$ to form $M$. 
In this limiting case, $D(x_F,M|s)$ should be independent 
of $\bar q_s(x_s)$ in the sense that if we further increase $\sqrt{s}$ 
we will gain nothing more. 
This implies that
$D(x_F,M|s)$ should be merely dependent 
on $q_v(x_F)$ but independent of $\bar q_s(x_s)$.
More precisely, at very high $\sqrt{s}$, 
we should have\cite{ft_lf}, 
\begin{equation}
D(x_F,M|s\to \infty)=\gamma_M q_v(x_F).
\end{equation}
where $\gamma_M$ is a constant. 
Hence, we have, 
\begin{equation}
A_N(x_F,s\to\infty)=
\frac{C\gamma_M \Delta q_v(x_F|tr)}{N_0(x_F|M)+\gamma_M q_v(x_F)},
\end{equation}
The constant $\gamma_M$ has a clear physical meaning: 
it is the probability for 
$q_v$ to hadronize into the meson $M$ 
in the quark-fusion process $q_v+\bar q_s\to M$. 
If we, for simplicity, take into account 
only vector and scalar meson production
with relative weights 3 to 1 due to spin counting, 
and take into account the suppression factor $\lambda$ (=0.3) 
for strange production, 
we have that, 
\begin{equation}
\gamma_\pi=1/[4(2+\lambda)].
\end{equation} 
We use this to do the calculations and obtain the results as shown 
by the dashed lines in Fig.2. 
We see that in this case $|A_N|$'s start to rise even 
slightly earlier than those at the E704 energy. 
This can be checked by future experiments.

In summary, we calculated the energy dependence of 
single spin asymmetries in inclusive pion 
production processes using the picture proposed in \cite{BLM93}. 
We showed that the energy dependence observed by BNL E925 
Collaboration compared to the Fermilab E704 results is a manifestation of the energy 
dependence of the interplay of the ``direct-formation'' part 
to the produced meson compared with the rest. 
We made predictions for experiments at event high energies. 

\vskip 0.2cm
This work was supported in part by the National Science Foundation
of China (NSFC) under the approval number 10175037 and
the Education Ministry of China
under Huo Ying-dong Foundation.

\begin{figure}
\includegraphics[]{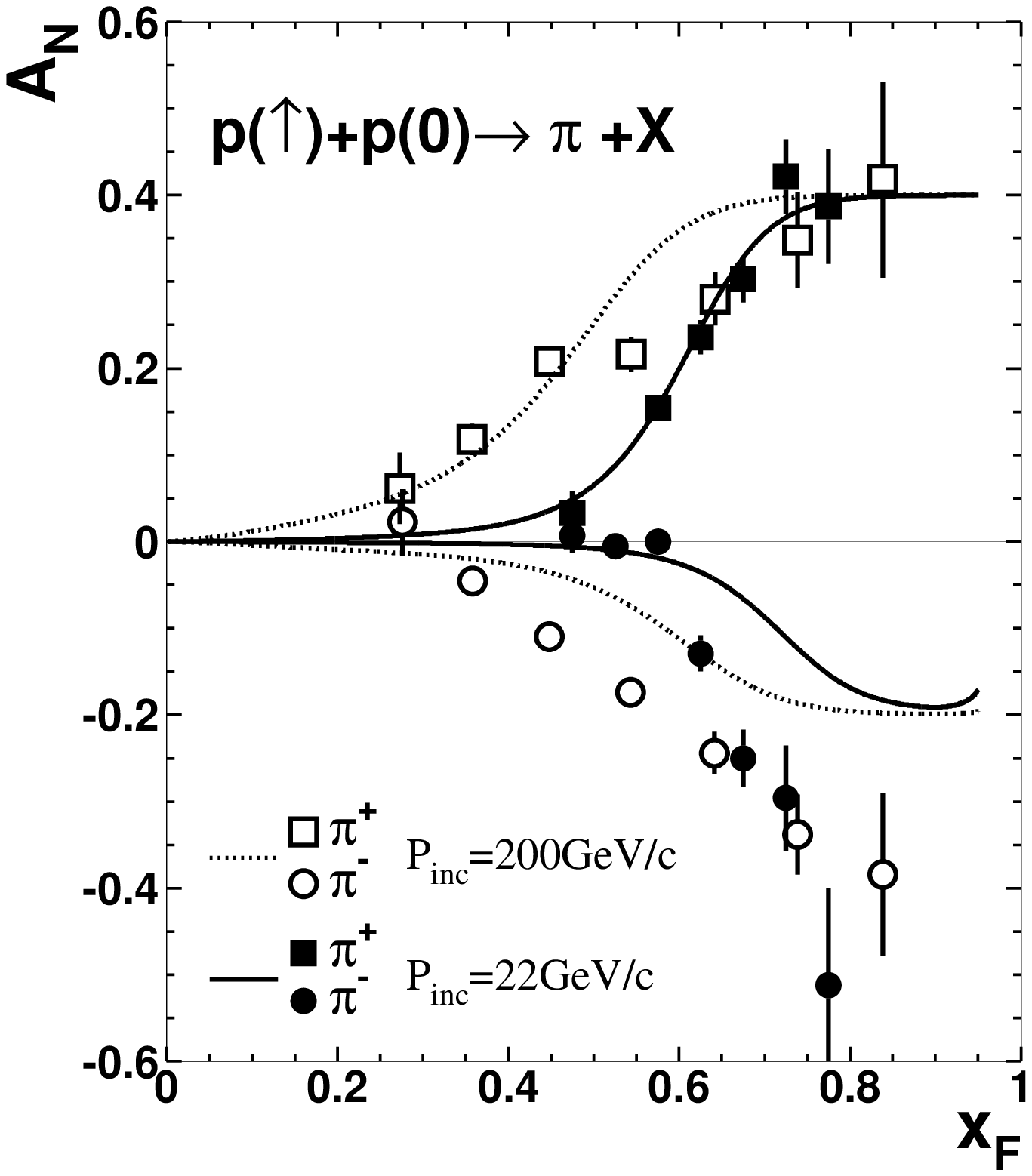}
\caption{\label{fig:AN93}
Single-spin asymmetry $A_N$ as a 
function of $x_F$ for $p(\uparrow)+
p(0)\rightarrow \pi^\pm+X$ at $p_{inc}$=200GeV/c 
compared with those at $p_{inc}$=22GeV/c . 
The data are taken from \cite{E704} and \cite{E925} respectively. 
The theoretical curves are the calculated results 
from Eq.(\ref{eq.define}) using the simple ansatz for 
$\Delta q_v(x|tr)\propto q_v(x)$ adopted in \cite{BLM93}.}
\end{figure}

\begin{figure}
\includegraphics[]{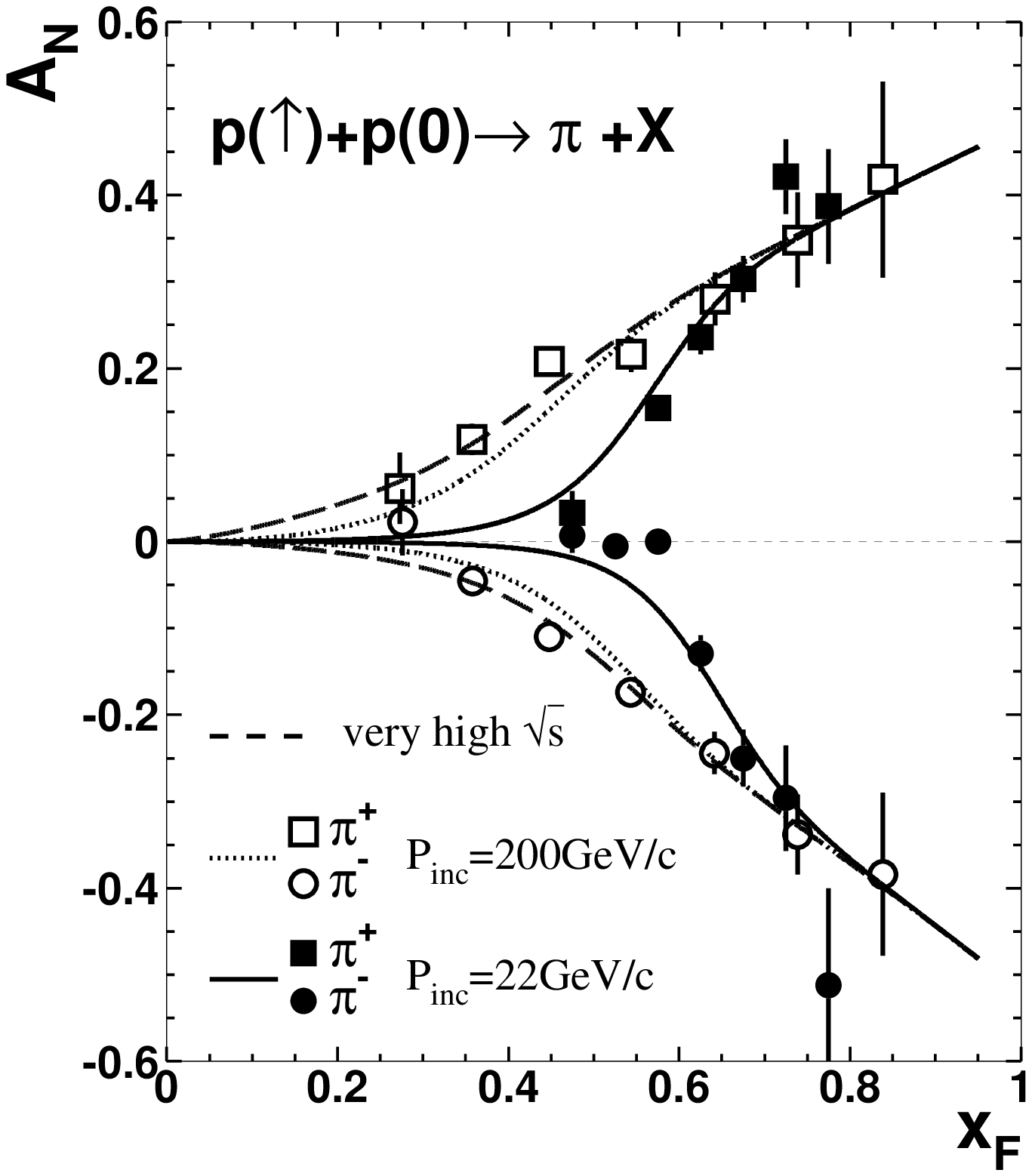}
\caption{\label{fig:AN03}
Single-spin asymmetry $A_N$ as a 
function of $x_F$ for $p(\uparrow)+
p(0)\rightarrow \pi^\pm+X$ at $p_{inc}$=200GeV/c 
(the two curves in the middle, dotted) 
compared with those at $p_{inc}$=22GeV/c 
(the two innermost curves, solid) and 
those at very high energy 
(the two outermost curves, dashed). 
The data are taken from \cite{E704} and \cite{E925} respectively. 
The curves are the calculated results 
from Eq.(\ref{eq.define}) using a modified parametrization of 
$\Delta q_v(x|tr)$.}
\end{figure}




\begin {thebibliography}{99}
\bibitem{Kane78} G.~Kane, J.~Pumplin and W.~Repko,
                 Phys. Rev. Lett. {\bf 41}, 1689 (1978).
\bibitem{ANexp} A summary of the data can be found in, e.g., 
  A.~Bravar, in proceedings of the 13th International Symposium 
  on High Energy Spin Physics, Protvino, Russia, ed. N.E.~Tyurin {\it et al}.,
  World Scientific (1999), p167; see in particular also 
  \cite{E704} listed below.
\bibitem{E704}  FNAL E704 Collab.,
      D.L.~Adams {\it et al}., Phys. Lett. B{\bf 264}, 462 (1991);
      B{\bf 276}, 531 (1992);
      Z. Phys. C{\bf 56},181 (1992).
      A.Bravar {\it et al}., Phys. Rev. Lett. {\bf 75}, 3073 (1995);
      {\bf 77}, 2626 (1996); 
      D.L.~Adams {\it et al.}, Nucl. Phys. B{\bf 510}, 3 (1998).
\bibitem{Siv90}  D. Sivers, Phys. Rev. D{\bf 41}, 83 (1990); 
                 D{\bf 43}, 261 (1991).
\bibitem{Qiu91}  J.~Qiu and G.~Sterman, Phys. Rev. Lett. {\bf 67}, 2264 (1991);
                 Phys. Rev. D{\bf 59}, 014004 (1999).
\bibitem{LM92}  Liang Zuo-tang and Meng Ta-chung,
                Z. Phys. A{\bf 344}, 171 (1992).
\bibitem{BLM93} C.~Boros, Liang Zuo-tang and Meng Ta-chung,
               Phys. Rev. Lett. {\bf 70}, 1751 (1993).
\bibitem{LM94} Liang Zuo-tang and Meng Ta-chung,
               Phys. Rev. D{\bf 49},3759 (1994); 
               C.~Boros, Liang Zuo-tang and Meng Ta-chung,
               Phys. Rev. D{\bf 51},4698 (1995); D{\bf 54}, 4680 (1996); 
              C.~Boros, and Liang Zuo-tang, Phys. Rev. D{\bf 53}, R2279 (1996).
\bibitem{Col93} J.~Collins, Nucl. Phys. B{\bf 394}, 169 (1993);
             and {\bf 396}, 161 (1993);
             J.~Collins, S.F.~Hepplelmann and G.A.~Ladinsky,
             Nucl. Phys. B{\bf 420}, 565 (1994).
\bibitem{Ans95} M.~Anselmino, M.~Boglione, and F.~Murgia,
             Phys. Lett.  B{\bf 362}, 164 (1995); 
             Phys. Rev. D{\bf 56}, 6021 (1997); {\bf 60}, 054027 (1999); 
             {\bf 67}, 074010 (2003).  
\bibitem{Efr95} A.V.~Efremov, V.M.~Korotkiyan, and O.V.~Teryaev,
             Phys. Lett. B{\bf 348}, 577 (1995).
\bibitem{Bro02} S. Bordsky, D.S. Hwang, I. Schmidt, 
            Phys. Lett. {\bf B530}, 99 (2002);
            Nucl. Phys. B{\bf 642}, 344 (2002);
            Int. J. Mod. Phys. {\bf A18}, 1327 (2003). 
\bibitem{Boer99} D. Boer, Phys. Rev. D{\bf 60}, 014012 (1999);
             D. Boer, and J.~Qiu, {\it ibid}, {\bf 65}, 034008 (2002).
\bibitem{Boer03} D. Boer, S. Brodsky, and D. Hwang, 
            Phys. Rev. D{\bf 67}, 054003(2003).
\bibitem{Bur02} M. Burkardt, Phys. Rev. D{\bf 66}, 114005 (2002).
\bibitem{LB00} For a review, see for example, 
            Liang Zuo-tang and C.~Boros, Inter. J. Mod. Phys. 
            {\bf A15}, 927 (2000).
\label{LB00}
\bibitem{E925} BNL E925 Collaboration,C. E. Allgower {\it et al.}, 
             Phys. Rev. D{\bf 65}, 092008 (2002).
\bibitem{ft_kappa} In practice, since there is no data 
directly on $N(x_F,s)$, $\kappa$ was obtained 
by fitting one data point at large $x_F$ of 
the available data on the invariant cross section 
$Ed^3\sigma/dp^3$ at given $p_\perp$ 
because approximately $Ed^3\sigma/dp^3\propto x_FN(x_F,s)$. 
More precisely, the number densities 
$N(x_F,s)$, $N_0(x_F,s)$ and $D(x_F,s)$ were all multiplied 
with $x_F$ and a constant which relates 
the differential cross section to the number density  
to become their counterparts in the invariant 
cross section $Ed^3\sigma/dp^3$. 
For $\pi$ at $p_\perp =0.65$ GeV/c, 
the corresponding result for $x_FN_0$ is 
$1.0e^{-2x_F^2-20x_F^3}$ mb/GeV$^2$ 
and that for $\kappa_\pi$ is $2\times 10^{-4}$ mb/GeV$^2$. 
See [7, 8, 16] for details.
\bibitem{Scaling} See, for example, 
            G.~Giacomelli and M.~Jacob, Phys. Rep. {\bf 55}, 38 (1979).
\bibitem{ft_pT} We neglected 
the influence of the transverse momentum $p_\perp$-dependence 
of $A_N$ in the calculations. 
This is because, as emphasized in \cite{E925}, the $p_\perp$ ranges in both experiments 
are approximately the same 
($0.2\le p_\perp \le 2$GeV/c for E704, and 
$0.3\le p_\perp \le 1.2$GeV/c for E925). 
On the other hand, it should be pointed out that 
the $p_\perp$ ranges in the two experiments are {\it not}
exactly the same. 
Hence, it can not be excluded that there might be some contribution 
from the $p_\perp$-dependence  of $A_N$
to the differences in the data obtained in the two experiments. 
This point should be further clarified by future experiments.   
\bibitem{ft_pdf} For the unpolarized quark distributions, we use the 
LO set (at $Q^2=1$GeV$^2$) of the GRV parametrization [M. Gl\"{u}ck, 
E. Reya, and A. Vogt, Eur. Phys. J. {\bf C5}, 461 (1998). ]. 
We should emphasize that the uncertainties 
at low $x$ region do not have much influence on our results.
This is because in the picture, the direct formation part dominates 
the large $x_F$ region. This is guaranteed by 
choosing an appropriate value for the constant $\kappa$ in the calculations.
\bibitem{ft_transversity} We take this form for simplicity and 
the obtained $\Delta q_v(x)$ satisfies the constraints from 
the general principles such as Soffer's positivity bound.
[J. Soffer, Phys. Rev. Lett. {\bf 74}, 1292 (1995).]  
\bibitem{ft_lf} This is consistent with the picture of limiting 
   fragmentation proposed in, 
   J. Benecke, T.T. Chou, C.N. Yang, and E. Yen, 
   Phys. Rev. {\bf 188}, 2159 (1969).  
\end{thebibliography}
\end{document}